\documentclass[twocolumn,amsmath,amssymb,preprintnumbers]{revtex4}
\usepackage{graphicx}% Include figure files
\usepackage{datetime}
\usepackage{hyperref,breakurl}
\renewcommand{\vec}[1]{{\bf #1}}
\newcommand{\exb}{{{\vec E}\!\times\!{\vec B}}}
\begin{document}
\title{Effect of turbulence on electron cyclotron current drive and heating in ITER}
\author{F.J.~Casson$^{1,2}$, E.~Poli$^1$, C.~Angioni$^1$, R.~Buchholz$^3$, A.G.~Peeters$^3$}
\address{$^1$ Max-Planck-Institut f\"{u}r Plasmaphysik, D-85748, Garching, Germany}
\address{$^2$ CCFE, Culham Science Centre, Abingdon, Oxon, OX14 3DB, UK}
\address{$^3$ Dept. of Physics, Universit\"at Bayreuth, D-95447, Bayreuth, Germany}

\begin{abstract}

Non-linear local electromagnetic gyrokinetic turbulence simulations of the ITER standard scenario H-mode are presented for the $q=3/2$ and $q=2$ surfaces.
The turbulent transport is examined in regions of velocity space characteristic of electrons heated by 
electron cyclotron waves.  Electromagnetic fluctuations and sub-dominant micro-tearing 
modes are found to contribute significantly to the transport of the accelerated electrons, even though they
have only a small impact on the transport of the bulk species.  The particle diffusivity for 
resonant passing electrons is found to be less than 0.15 $m^2 s^{-1}$, 
and their heat conductivity is found to be less than 2 $m^2 s^{-1}$.
Implications for the broadening of the current drive and energy deposition in ITER are discussed. 

\end{abstract}

\maketitle

%\section{Introduction}

In ITER, externally applied millimeter waves will be used to heat and drive current
in the plasma through resonances with the electron cyclotron frequency.
The deposition width of Electron Cyclotron Current Drive and Heating (ECCD and ECH respectively)
is usually calculated using ray and beam tracing codes neglecting the effects of plasma turbulence. 
The turbulence can affect the deposited energy in two ways: 
either by scattering the incoming waves before they are absorbed \cite{tsironis_electron-cyclotron_2009,weber_investigation_2013}, or by 
transporting the heated electrons before they have interacted with the bulk plasma, causing spreading or shifting 
of the deposited energy \cite{peeters_impact_1996,kirov_ecrh_2002} or driven current \cite{harvey_radial_2002}. 
In this work, we examine the latter effect for the parameters of the ITER baseline H-mode scenario, at the $q=3/2$ and $q=2$ rational flux surfaces.

In Ref. \cite{bertelli_consequences_2009}, Fokker-Planck calculations of EC wave absorption in ITER
including an arbitrarily prescribed turbulent diffusivity for the heated electrons demonstrated that a particle diffusion of the order of 1 $m^2 s^{-1}$ could broaden the current deposition profile sufficiently to pose difficulties for Neoclassical Tearing Mode (NTM) stabilisation.  Here, we calculate these diffusion coefficients, using self-consistent electromagnetic local gyrokinetic simulations with the GKW code \cite{peeters_nonlinear_2009}. 
The simulations in the present work do not include any large scale magnetic island, which is known to flatten the profiles and reduce the turbulence inside the island \cite{poli_behaviour_2009,hornsby_nonlinear_2010,hornsby_nonlinear_2014}.  As such, the results in this work represent upper limits for the turbulent diffusion of heated electrons that can be expected in ITER, such as in the case of pre-emptive NTM stabilisation.

%\section{Simulation parameters}

The physical parameters for the simulations were obtained from scenario modelling \cite{parail_self-consistent_2013}\footnote{The flat top phase of Case\#001 from Ref. \cite{parail_self-consistent_2013} is used, which corresponds to the run with PPF file\newline  \texttt{\#53287/fkochl/jul2811/seq.1/ppfseq.17416}}, with the JINTRAC code \cite{romanelli_jintrac:_2014}, and are given in Table I.  
Using the JINTRAC parameters (later called `nominal'), which predicts turbulent heat fluxes using the GLF23 code \cite{waltz_gyro-Landau-fluid_1997},
the non-linear gyrokinetic simulations find heat losses that are too high to be compatible with the global scenario. The GLF23 model uses a simplified 
geometry which is known to underpredict the fluxes \cite{kinsey_iter_2011}, as a result it is likely that it has optimistic temperature profiles.  The temperature gradients were therefore reduced in our simulations, to give more realistic (but still high) turbulent total power losses $P_{i+e}$. Due to the inherent gyro-Bohm scaling of the gyrokinetic results, gradient driven, flux matched simulations of ITER will always sit close to marginal stability, so attempting to more closely match predicted fluxes to the scenario is impractical.  With the reduced gradients, the electron conductivities are close to the scenario values;   
it is not possible to match both ion and electron conductivities to the JINTRAC values, because the ITG dominated nature of the turbulence
gives ion heat flux 2-3 times larger than the electron heat flux in all cases. %(and ETG fluxes are expected to make only a small contribution \cite{kinsey_iter_2011} - DO WE NEED TO CHECK THIS FOR THE FAST ELECTRONS?).  
Given these difficulties with flux matching, 
the diffusivities presented are normalised to the total heat conductivity $\chi_{i+e} = -( Q_i + Q_e )/(n_i\nabla T_i + n_e\nabla T_e)$
of the simulation. To obtain dimensional diffusivities with realistic values useful for the Fokker-Planck calculations, the dimensionless ratio is multiplied by the total heat conductivity from the scenario modelling, which matches realistic heat fluxes.  By varying the temperature gradients between the nominal and reduced values, we have verified that these dimensionless ratios are insensitive to changes in the total fluxes. 

For this work, it is important that the particle fluxes be consistent with the steady state scenario with relatively small core fuelling.  The JINTRAC scenario uses a prescribed density profile, which the gyrokinetic simulations found to be somewhat pessimistic, with a strong inward particle flux.  The density gradient was therefore increased from the nominal value (Table 1), to give a near zero particle flux ($\Gamma T_i / Q_i < 0.03$), i.e. a steady state with negligible sources. 

\begin{table*}[t!]
\squeezetable
\begin{footnotesize}
\begin{tabular}{|l|l|l|l|l|l|l|l|l|l|l||l|l|l|l|}
\hline
$q$  & $ {\hat s}$ & $r/R$ & $R/L_{T_e}$ & $ R/L_{T_i} $ & $R/L_n$ & $\beta_e $ & $Z_{\rm eff}$  &  $T_e$(keV)   & $T_i$ (keV)   & $n_e (m^{-3})$  &  $\chi_i (m^2/s)$   &  $\chi_e (m^2/s)$ & $\chi_{i+e} (m^2/s)$ & $ P_{i+e}$(\textsc{mw}) \\ %& $\Delta t (\frac{v_{\rm th}}{R})$\\
\hline
3/2  &   1.63       &  0.230   & 5.00*         &   4.90*         &  1.50*   & 1.22\%     & 1.76   &   8.77          & 8.57          &  9.72 $\cdot 10^{19}$                  &  \textit{0.89} (\texttt{2.74})     &   \textit{0.93} (\texttt{0.65})&  \textit{0.91} (\texttt{1.69}) &  $\sim$\textit{80} (\texttt{135}) \\ %& 339 \\
  2  &   2.39       &  0.267  & 5.00*         &   4.90*         &  1.50*   & 0.97\%     & 1.76   &   7.06          & 6.78          &  9.56 $\cdot 10^{19}$                &  \textit{1.01} (\texttt{5.74})      &   \textit{1.03} (\texttt{1.45}) & \textit{1.02} (\texttt{3.58}) & $\sim$\textit{80} (\texttt{263}) \\ %&  355 \\
\hline
\end{tabular}
\end{footnotesize}
\caption{Physical input parameters for the simulations from the JINTRAC scenario (*= Values adapted; unmodified values 
$(R/L_{T_e},~R/L_{T_i},~R/L_n) = (5.60, 5.73, 0.43),~(5.95, 6.33, 0.49)$ for the $q=3/2$ and $q=2$ cases respectively).  The turbulent heat conductivities $\chi_{i,e,i+e}$ and power loss $P_{i+e}$ are simulation outputs for comparison; values in \textit{italic} indicate scenario values, values in \texttt{typed font} are those in the GK simulation.} 
\end{table*}

The simulations include electromagnetic fluctuations in $A_\parallel$, linearised 
pitch-angle collisions between all species including a factor $Z_{\rm eff} = 1.76$ for electron-ion collisions,
and full flux-surface geometry from the scenario modelling.  
A single ion species is used, with a mass equivalent to a DT 50:50 mixture, with two kinetic electron species (explained below).
The velocity grids were extended beyond their default settings to $(v_\parallel^{\rm max},v_\perp^{\rm max})\,=\,(4,4)v_{\rm th}$, to capture the velocity space of interest for the ECCD resonance (in this work, $v_\perp = \sqrt{2 \mu B_A/m}$, where $B_A$ is the magnetic field at the axis $R = 6.2m$, and $v_{\rm th} = \sqrt{2T/m}$, and $\rho_i = m v_{th} / e B_A$).  The velocity grid has $48 \times 16$ points in $v_\parallel,\mu$, and the parallel grid has 36 points \footnote{Input files for the presented simulations are archived publicly at \newline \url{http://gkw.googlecode.com/svn/input/2014_Casson_NF}; the GKW source is also obtainable from this repository.}.  

%\subsection{Unstable eigenmodes}

\begin{figure}[tbh!]
\centering
\includegraphics[width=72mm]{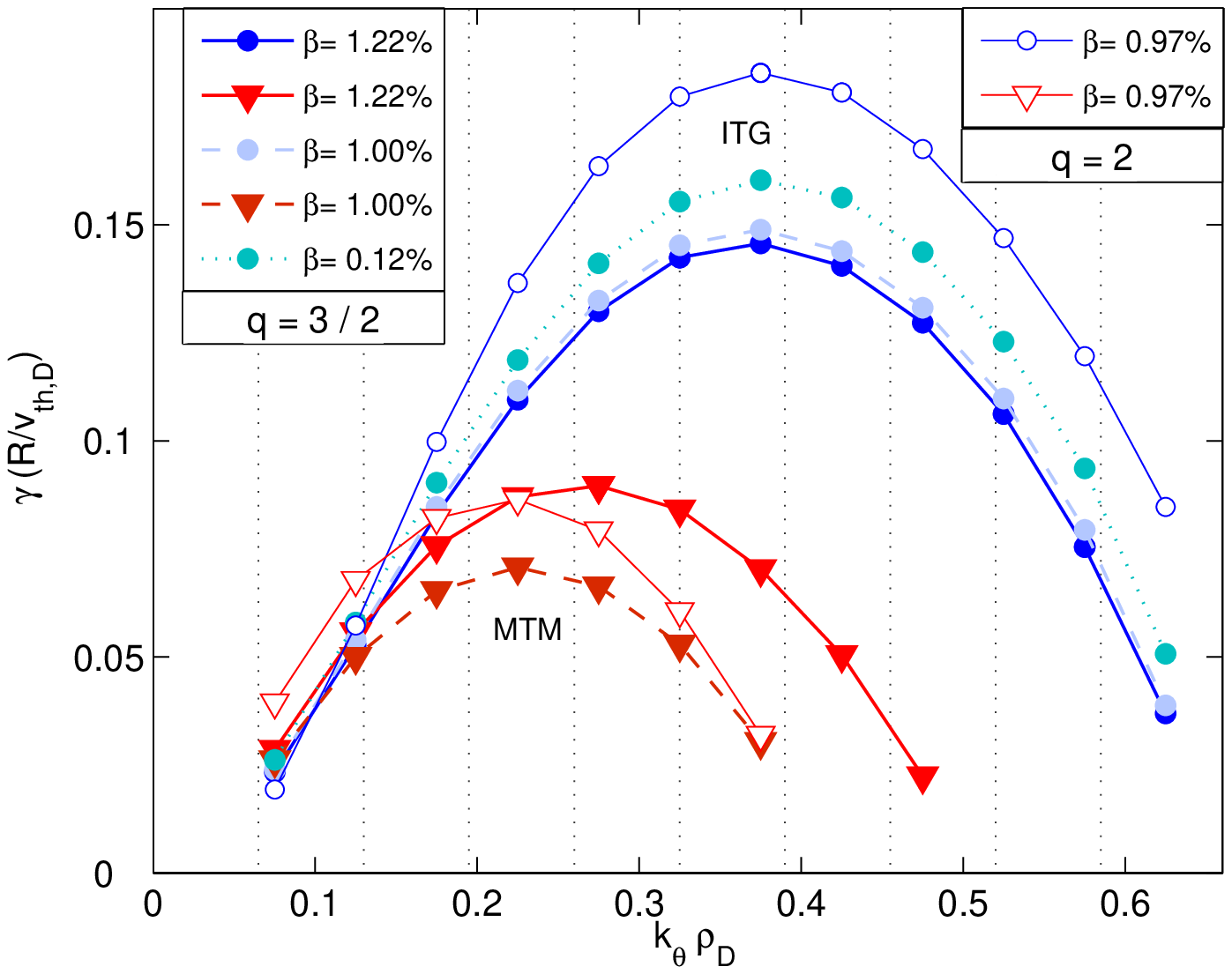}
\includegraphics[width=72mm]{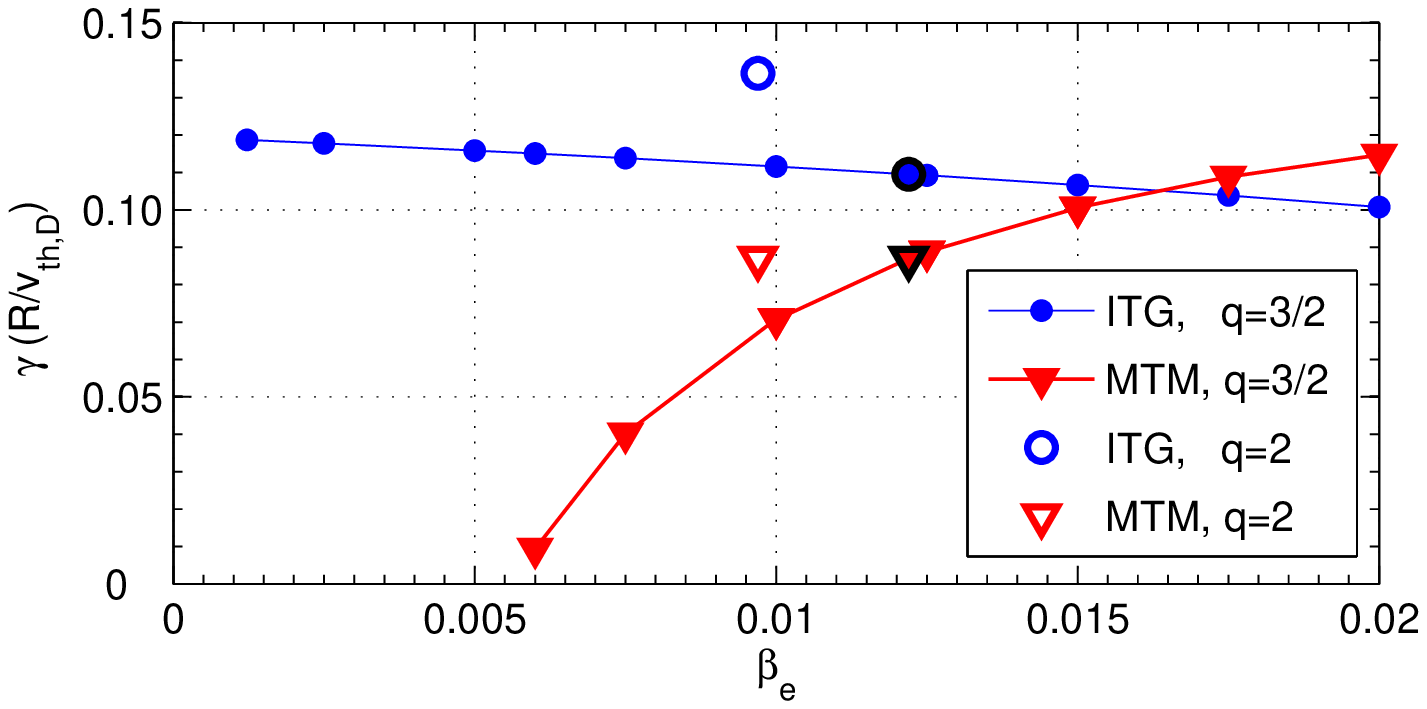}
\caption{(top) Eigenmode growth rate spectra at the $q=3/2$ surface (filled symbols) for $\beta_e = \{0.122\%,~1.00\%,~1.22\%\}$, and for $q=2$ (open symbols) with nominal $\beta_e=0.97\%$.  The vertical dotted lines indicate the lowest modes in the non-linear simulation.
(bottom) Growth rates at $k_\theta \rho_D = 0.225$ for a $\beta_e$ scan. 
For nominal $\beta_e$ (thick symbols) the ITG mode (red circles) is dominant; a sub-dominant MTM with tearing parity (blue triangles) is also present for all $\beta_e > 0.06\%$.}
\label{fig.beta_scan_lin}
\end{figure}

The eigenmode stability is investigated using a newly implemented eigenvalue solver in GKW,
which exploits a matrix-free method of the SLEPc library \cite{hernandez_slepc:_2005}.  The results, in Fig. \ref{fig.beta_scan_lin}, find the expected dominant ITG instability at all scales $k_\theta \rho_D < 0.7$, and a sub-dominant micro-tearing (MTM) instability at slightly lower $k_\theta \rho_D < 0.5 $ (similar to Ref. \cite{saarelma_mhd_2013}).  Given the sensitivity of the MTM to $\beta_e$ \cite{gladd_electron_1980,applegate_micro-tearing_2007}, we also performed a $\beta_e$ scan for the $q=3/2$ case, which demonstrates that moderate increases in $\beta_e$ could change the dominant instability to MTM.  At higher wavenumbers (not shown), there are well-separated ETG modes with growth rates (normalised to $v_{\rm th,e}$) less than the MTM, consistent with the ITER expectation \cite{kinsey_iter_2011} that electron scales will contribute only a small part to the transport.

In the non-linear simulations, $21 \times 167$ Fourier modes are used, %in the bi-normal and radial directions respectively, 
with maximum $k^{\rm max}_\theta \rho_D = 1.3$, and a low field side perpendicular box size $[L_x,L_y]=[40,97]\rho_D$.  The portion of the spectrum unstable to micro-tearing is covered by 7 bi-normal modes (Fig. \ref{fig.beta_scan_lin}).  Convergence tests in $k^{\rm min}_\theta$, $k^{\rm max}_\theta$ and parallel resolution did not change the results by more than $10\%$.  The sensitivity of both linear and non-linear results to the full linearised Landau-Boltzmann collision operator was also tested; no appreciable difference was found. The presented results are all time-averaged for $\Delta t \sim 350 R/v_{\rm th,D}$ after non-linear saturation is reached.  The integrated fluxes show the characteristics expected for electromagnetic ITG turbulence: Increasing $\beta_e$ stabilises the turbulence, such that at the nominal $\beta^{\rm ITER}_e$, the simulation is only marginally unstable. The electromagnetic flutter heat flux is inwards up to $k_\theta \rho_D = 0.4$ (which indicates that ITG modes are strongly dominant over MTM at these scales \cite{doerk_gyrokinetic_2012,hatch_magnetic_2013}), but 20 times smaller (in magnitude) than the $\exb$ flux.  In the particle transport channel, the total magnetic flutter particle fluxes are also small.  The convective particle flux $RV_{E \times B}$ does not change significantly with $\beta_e$, but the diffusive flux increases outwards as $\beta_e$ increases.

%\section{Velocity decomposed fluxes}
To extract diffusivities for the electrons accelerated by EC resonances, we assume these electrons exist in trace concentration and respond to (but do not modify) any turbulence generated by the bulk species.  For trace species, it was shown in Ref. \cite{angioni_gyrokinetic_2008} that the velocity structure of the background does not affect the kernel of the fluxes, i.e., that the flux for an alternative distribution can be accurately computed by integrating the flux kernel with the appropriate background distribution.  For this study, fluxes are output from the code without any velocity space integration, and we define the velocity decomposed flux $\Gamma^\vec{v}$ for a species $s$ such that 
\begin{equation}
\Gamma_s = \biggl \langle \int [\tilde {\bf v}_E, {\bf v}_{\delta B}]  \cdot \nabla \psi f {\rm d}^3 {\bf v} \biggr \rangle \equiv \int \Gamma^\vec{v}_s d^3\vec{v}
%+ \biggl \{ \int {\rm d}^3 {\bf v} \tilde {\bf v}_{\delta B}\cdot \nabla \psi  \frac{mv^2}{2} f \biggr \} 
\end{equation}
is the total flux (with similar notation for $D^\vec{v}, \chi^\vec{v}$, etc).  The first term in the $[,]$ brackets gives the transport due to the perturbed electric field, the second gives the transport due to the magnetic flutter (respectively denoted $\exb$ and MF hereafter). The flux surface average $\langle \rangle$ commutes with the velocity integral, therefore the $\Gamma^\vec{v}_s$ are flux surface averaged.

\begin{figure}[t]
\includegraphics[width=80mm]{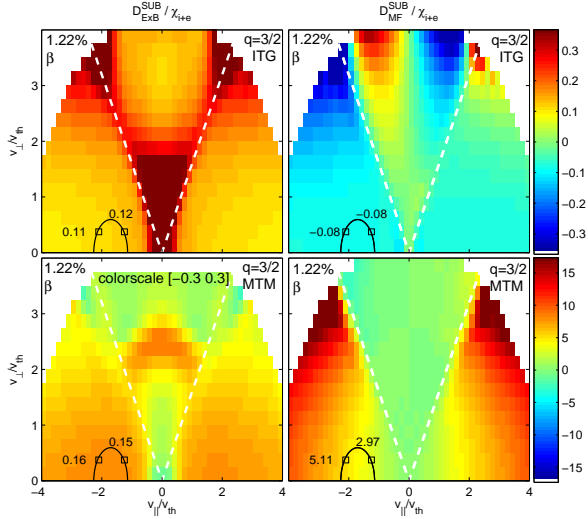}
\caption{Velocity kernel of electron diffusivity for the two linear eigenmodes (top: ITG, dominant; bottom: MTM, sub-dominant) on the $q=3/2$ surface at nominal $\beta_e = 1.22\%$ and $k_\theta \rho_D = 0.225$.
%The electrostatic ${\tilde v}_E$ and electromagnetic flutter ${\tilde v}_{\delta B}$ radial transport channels are marked with ES and EM respectively.  Numbers in the plot indicate the values at selected points on the ECCD resonance (marked by x).  
 The dashed white line shows the trapped-passing boundary, and the black line is the EC resonance.  The two top plots share the same colorscale.  In the bottom left the range of the colorscale is 1/50 that of the bottom right.}
\label{fig.Deiv}
\end{figure}

\begin{figure}[t]
\includegraphics[width=80mm]{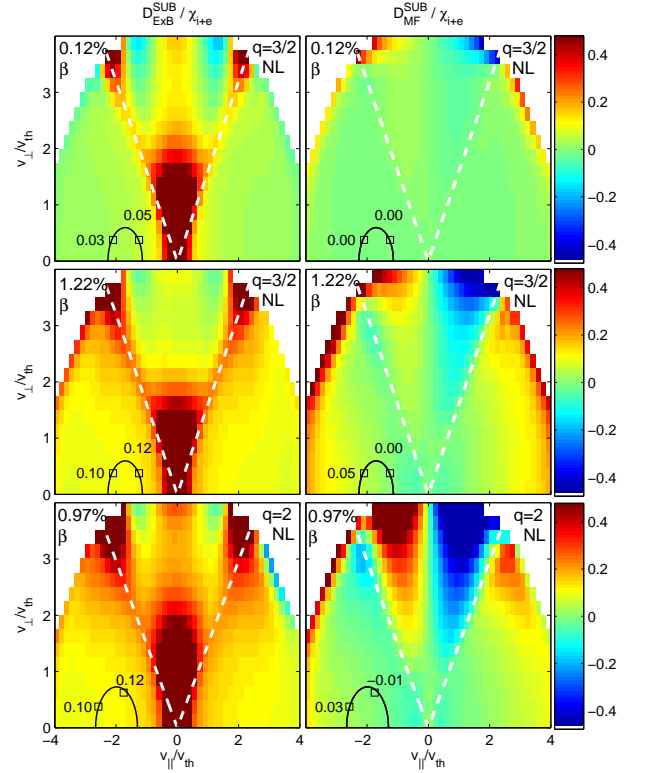}
\caption{Velocity kernel of electron diffusivity, for the near electrostatic ($q = 3/2,~\beta_e = 0.122\%$, [top]) and full electromagnetic non-linear simulations ($q = 3/2,~\beta_e = 1.22\%$ [mid], $q = 2,~\beta_e = 0.97\%$ [bottom]).}
%The electrostatic ${\tilde v}_E$ and electromagnetic flutter ${\tilde v}_{\delta B}$ radial transport channels are marked with ES and EM respectively.  Numbers in the plot indicate the values at selected points on the ECCD resonance (marked by x).
\label{fig.Dturb}
\end{figure}

\begin{figure}[]
\includegraphics[width=80mm]{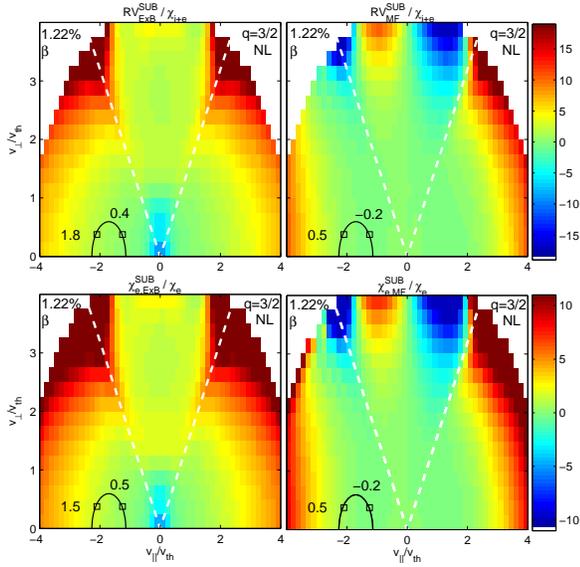}
\caption{Velocity kernel of electron convection (top) and heat conductivity (bottom), for the full electromagnetic non-linear simulation at the $q=3/2$ surface.\label{fig.Chiturb}}
\end{figure}

The electron flux $\Gamma_e$ is decomposed into diffusive and convective components respectively
\begin{equation}
 \frac{R \Gamma_e}{n} = - D \frac{R}{L_n} + R V
\end{equation}
by the use of an additional trace electron species with ${R}/{L_{n_e}}=0$ (we have verified that using two non-trace electron species 
with the same total $R/L_{n_e}$ gives identical results). The velocity decomposition of the fluxes are retained, 
so that from the velocity decomposed diffusivity (and similarly for convection), we define the velocity kernel of the diffusivity
\begin{equation}
D_{\rm Sub} = \frac{\int_{\rm Tot} F_M {\rm d}^3 {\vec v} }{\int_{\rm Sub} F_M {\rm d}^3 \vec{v} }\cdot \int_{\rm Sub} D^\vec{v} {\rm d}^3 \vec{v}.
\end{equation}
which is the contribution of a sub-region of velocity space to diffusion, normalised to the density in that region (e.g. the diffusivity at a given velocity).  The velocity kernel of the heat conductivity, $\chi_e^{\rm Sub}$, is similarly defined using $(\int_{\rm Tot} v^2 F_M {\rm d}^3 {\vec v} )/(\int_{\rm Sub} v^2 F_M {\rm d}^3  \vec{v})$ as the dimensionless normalising factor.  Towards the edge of velocity space, contributions to the total flux are small, but the local density (the denominator) is also small; regions in which the values are too small 
for an accurate machine representation are excluded (in white) from Figs. \ref{fig.Deiv}-\ref{fig.Chiturb}.  Because of this denominator, values increasing towards the edge of the domain do not indicate an under-resolved simulation, but spreading of the perturbation beyond the background Maxwellian.  The values in the figures can be multiplied with the scenario $\chi_{i+e}$ given in Table I, and integrated over the fast electron distribution output by a Fokker-Planck solver to determine total diffusivities for the heated electrons.  For an order-of-magnitude estimate to compare with the critical $D \sim 1m^2s^{-1}$ in Ref. \cite{bertelli_consequences_2009}, we here analyse specific points (black squares) in the region of the EC resonance (black curve) for the ITER NTM stabilisation scheme.  The resonance is calculated from the relativistic cyclotron resonance condition (for absorption at the fundamental harmonic) $\omega-\Omega/\gamma-k_\parallel v_\parallel=0$, where $\omega$ is the angular wave frequency, $\Omega$ is the cyclotron frequency, $\gamma$ is the relativistic Lorentz factor and $k_\parallel$ and $v_\parallel$ are the parallel components of the wavevector and particle velocity.  The parameters are evaluated at the position correponding to the maximum of the absorption profile according to the beam tracing code TORBEAM \cite{poli_torbeam_2001}.

%\begin{equation}
%\chi_{\rm Sub} = \int_{\rm Sub} \chi^\vec{v} {\rm d}^3 \vec{v} \cdot \frac{\int_{\rm Tot} v^2 F_M {\rm d}^3 {\vec v} }{\int_{\rm Sub} v^2 F_M {\rm d}^3 %\vec{v} }
%\end{equation}

Comparing the velocity kernel of the particle diffusivity between the $q=3/2$ eigenmodes (Fig. \ref{fig.Deiv}) and the non-linear simulations, 
%at nominal $\beta^{\rm ITER}_e$ and reduced $\beta_e = \beta^{\rm ITER}_e / 10$ (Fig. \ref{fig.Dturb}) 
we draw a number of conclusions:  The values and shape of $D^{\rm Sub}_\exb/\chi_{i+e}$ for the ITG eigenmode is similar to the full electromagnetic non-linear simulation, since ITG modes dominate the turbulence.   However, the diffusivity $D^{\rm Sub}_{\rm MF}$ for the ITG mode is always inwards in the passing domain, but is strongly outwards for the MTM.  In the non-linear simulation, the passing domain $D^{\rm Sub}_{\rm MF}$ is outwards, which indicates that the sub-dominant MTM plays a dominant role here, even though its effect on the integrated fluxes is small.  We note also that the $D^{\rm Sub}_\exb$ for the ITG has an outward component in the passing domain, which increases with $\beta_e$.  We thus conclude that the $D^{\rm Sub}_\exb$ channel is everywhere dominated by ITG modes, but the $D^{\rm Sub}_{\rm MF}$ channel is dominated by sub-dominant MTM in the passing domain. For the $q=3/2$ case, the total diffusivities are $D^{\rm Sub} \sim \{0.14,0.11\}~m^2 s^{-1}$ for the $v_\parallel = \{-2.1,-1.25\} v_{\rm th}$ points respectively ($\{,\}$ notation repeated below).
 
As expected from the linear eigenvalues, the $q=2$ surface shows very similar results to $q=3/2$ (with diffusivities $D^{\rm Sub} \sim \{0.13,0.11\}~m^2 s^{-1}$), but due to its lower $\beta_e$, the transport in the $D^{\rm Sub}_{\rm MF}$ channel is reduced.  The MTM stability is here not strongly affected by the geometry.  There are no kinetic ballooning modes present in these simulations, so no scaling of electromagnetic effects with $q^2$ should be expected (and none is evident). 

The influence of the sub-dominant MTM is also present in the kernel of the heat conductivity (Fig. \ref{fig.Chiturb}, bottom), which also mirrors the form and magnitude of the ITG eigenmode in the $\chi^{\rm Sub}_{e,\exb}$ channel, and the form (but not magnitude) of the MTM eigenfunction in the $\chi^{\rm Sub}_{e,\rm MF}$ channel. The sub-dominant MTM again contributes significantly to the conductivity $\chi^{\rm Sub}_{e,\rm MF}$ in the passing domain, and the electromagnetic fluctuations on the ITG increase the $\chi^{\rm Sub}_{e,\exb}$ contribution.  Together, for the $q=3/2$ case, these contributions give $\chi^{\rm Sub}_e \sim \{1.9,0.3\}~m^2 s^{-1}$. 
This heat conductivity could be used to determine the broadening of the region in which the heated electrons equilibrate with the surrounding plasma. 
However, 
since this profile establishes on a time of the order of the electron transit time (which is much faster than the 
transport time scales), this broadening will be neglible.

The passing particles also exhibit a significant convection (Fig. \ref{fig.Chiturb}, top), with values and functional form similar to the heat conductivities $\chi_e^{\rm Sub}/\chi_{i+e}$.  %For the $v_\parallel = \{2.25,3.25\} v_{\rm th}$ points,
The values for the $q=3/2$ case have total $RV^{\rm Sub} \sim \{2.1,0.2\} ~m^2 s^{-1}$.  The similarity to the heat conductivity indicates that the convection is thermo-diffusive, and primarily driven by $R/L_{T_e}$ \cite{angioni_particle_2009}.  A large convection could
also affect the current deposition profile by shifting the accelerated electrons.  A simple estimate for a full deposition width $w_{\rm cd} \sim 4$ cm gives convective timescales $\tau_{\rm conv} \approx w_{\rm cd}/2V \approx \{0.057,0.600\}s$ substantially slower the diffusive timescales $\tau_{\rm D} \approx w_{\rm cd}^2/4D \approx \{0.0029,0.0036\}s$, which indicates that the convective effect may be neglected (at least for this narrow $w_{\rm cd}$).  

\begin{figure}[t!]
\includegraphics[width=55mm,bb=158 164 452 627]{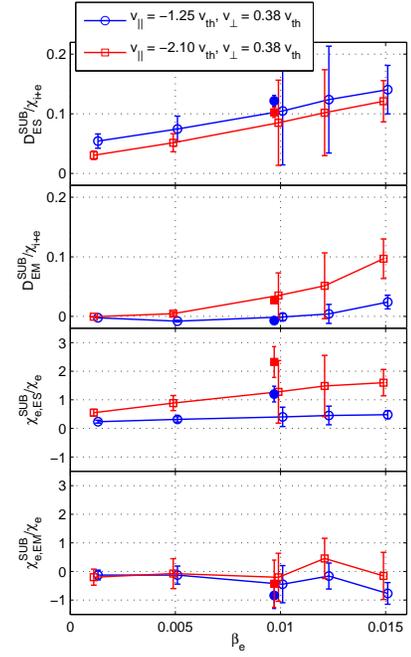}
\caption{Transport coefficients for the points in Figs. \ref{fig.Dturb} and \ref{fig.Chiturb} as a function of $\beta_e$, for both the $q=3/2$ (open) and $q=2$ (filled) surfaces.  The error bars are the standard deviation of four distinct ranges of the time-average after saturation.}
\label{fig.beta_scan_nl}
\end{figure}

Given the sensitivity of the MTM to $\beta_e$, and its stability threshold (Fig. \ref{fig.beta_scan_lin}), we examine the diffusion and conductivity at selected velocity points in a $\beta_e$ scan (Fig. \ref{fig.beta_scan_nl}).  The $D^{\rm Sub}_\exb$ and $\chi^{\rm Sub}_\exb$ components scale linearly with $\beta_e$, consistent with the hypothesis that they are driven by electromagnetic perturbations to ITG modes, while the $D^{\rm Sub}_{\rm MF}$ and  components increase stronger than linearly ($\sim \beta_e^2$) above $\beta_e = 0.005$, consistent with the stability threshold and transport scaling of the MTM \cite{doerk_gyrokinetic_2012,hatch_magnetic_2013}.

%The $\chi_e$ and $RV$ components show the same structure in velocity space, and the same scaling with $\beta_e$, which indicates that they are both due to the same %$R/L_{T_e}$ driven population of the perturbed distribution.

To conclude: In this work we have used electromagnetic gyrokinetic simulations to examine the ion-scale turbulence on the $q=3/2$ and $q=2$ rational flux surfaces of the ITER baseline H-mode scenario, which are of interest for NTM stabilisation by ECCD.  We find that sub-dominant micro-tearing modes can significantly enhance passing electron transport, even though their contribution to the integrated fluxes of the bulk species is small.  The fast electron transport is therefore sensitive to the value of $\beta_e$, but at the scenario reference values of $\beta_e$, the turbulent transport of accelerated electrons is insufficient to cause a significant spreading of the current deposition profile ($D \sim 0.15 m^2s^{-1}$, less than the critical $D \sim 1m^2s^{-1}$ in Ref. \cite{bertelli_consequences_2009}), and should not pose problems for NTM stabilisation.  
Future work should verify this for electron-scale turbulence, potentially more important in advanced scenarios.  
The macroscopic magnetic island was not included in these simulations, but the reduction of turbulence due to an island can only reinforce this conclusion.  

\begin{acknowledgments}
FJC would like to thank G. Colyer, H. Doerk, E. Fable, T. G{\"o}rler, F. K{\"o}chel, O. Maj, C.M. Roach, and H. Weber, 
for useful comments and discussions.  Simulations were performed using the Helios supercomputer at IFERC-CSC, 
and the resources of the Rechnungzentrum at Garching.
This work has received funding from the European Union's Horizon 2020 research and innovation programme
under grant agreement number 633053, the RCUK Energy Programme [grant number EP/I501045], and the Max Planck Institute.
\end{acknowledgments}

\bibliographystyle{my_doi2}
\bibliography{iter_paper}

%\begin{thebibliography}{10}
%\end{thebibliography}

\end{document}